# ULTRASONIC PRODUCTION OF NANO-SIZE DISPERSIONS AND EMULSIONS

*Thomas Hielscher*

Dr. Hielscher GmbH, Warthestrasse 21, 14513 Teltow, Germany, (http://www.hielscher.com)

**ABSTRACT**

Ultrasound is a well-established method for particle size reduction in dispersions and emulsions. Ultrasonic processors are used in the generation of nano-size material slurries, dispersions and emulsions because of the potential in the deagglomeration and the reduction of primaries. These are the mechanical effects of ultrasonic cavitation. Ultrasound can also be used to influence chemical reactions by the cavitation energy. This is sonochemistry. As the market for nano-size materials grows, the demand for ultrasonic processes at production level increases. At this stage, energy efficiency becomes important. Since the energy required per weight or volume of processed material links directly to the equipment size required, optimization of the process efficiency is essential to reduce investment and operational costs. Furthermore it is required to scale the lab and bench top configurations to this final level without any variations in the process achievements. Scale up by power alone will not do this.

## 1. INTRODUCTION

Ultrasound is a very effective processing method in the generation and application of nano-size materials. In general, ultrasonic cavitation in liquids may cause fast and complete degassing: initiate various chemical reactions by generating free chemical ions (radicals); accelerate chemical reactions by facilitating the mixing of reactants; enhance polymerization and depolymerization reactions by temporarily dispersing aggregates or by permanently breaking chemical bonds in polymeric chains; increase emulsification rates; improve diffusion rates; produce highly concentrated emulsions or uniform dispersions of micron-size or nano-size materials; assist the extraction of substances such as enzymes from animal, plant, yeast, or bacterial cells; remove viruses from infected tissue; and finally, erode and break down susceptible particles, including micro-organisms [2]. Ultrasound can be tested in lab and bench-top scale before the results are scaled up to the commercial level.

## 2. ULTRASONIC CAVITATION

Low-intensity or high-frequency ultrasound is mainly used for analysis, non-destructive testing and imaging. High-intensity ultrasound is used for the processing of liquids such as mixing, emulsifying, dispersing and deagglomeration, or milling. When sonicating liquids at high intensities, the sound waves that propagate into the liquid media result in alternating high-pressure (compression) and low-pressure (rarefaction) cycles, with rates depending on the frequency. During the low-pressure cycle, high-intensity ultrasonic waves create small vacuum bubbles or voids in the liquid. When the bubbles attain a volume at which they can no longer absorb energy, they collapse violently during a high-pressure cycle. This phenomenon is termed cavitation. Cavitation, that is "the formation, growth, and implosive collapse of bubbles in a liquid. Cavitational collapse produces intense local heating (~5000 K), high pressures (~1000 atm), and enormous heating and cooling rates ($>10^9$ K/sec)" and liquid jet streams (~400 km/h)" [4]. Cavitation can be produced in different ways: e.g. by high-pressure nozzles, rotor-stator mixers, or ultrasonic processors. In all those systems the input energy is transformed into friction, turbulences, waves and cavitation. The fraction of the input energy that is transformed into cavitation depends on several factors describing the movement of the cavitation generating equipment in the liquid. The intensity of acceleration is one of the most important factors influencing the efficient transformation of energy into cavitation. Higher acceleration creates higher-pressure differences. This in turn increases the probability of the creation of vacuum bubbles instead of the creation of waves propagating through the liquid. Thus, the higher the acceleration the higher is the fraction of the energy that is transformed into cavitation. In case of an ultrasonic transducer, the amplitude of oscillation describes the intensity of acceleration. Higher amplitudes result in a more effective creation of cavitation. In addition to the intensity, the liquid should be accelerated in a way to create minimal losses in terms of turbulences, friction and wave generation. For this, the optimal way is a unilateral direction of movement.

## 3. MATERIALS

Nanomaterials have become integral components of products as diverse as sunscreens, electrically conductive coatings, and strong, lightweight plastic composites or paints and inks. The production and processing of these materials is a quickly progressing technology. Indeed, it is advancing rapidly and creating new possibilities for many industrial processes. Solid nanomaterials fall into three broad categories: metal oxides, nanoclays, and carbon nanotubes. Metal-oxide nanoparticles include nanoscale zinc oxide, titanium oxide, iron oxide, cerium oxide and zirconium oxide, as well as mixed-metal compounds such as indium-tin. In general, nanomaterials are defined as materials of less than 100nm in size. When matter is reduced in size it changes its characteristics, such as color and interaction with other matter such as chemical reactivity. The change in the characteristics is caused by the change of the electronic properties. By the particle size reduction, the surface area of the material is increased. Due to this, a higher percentage of the atoms can interact with other matter, e.g. with the matrix of resins. Surface activity is a key aspect of nanomaterials. Agglomeration and aggregation blocks surface area from contact with other matter. Only well-dispersed or single-dispersed particles allow to utilize the full beneficial potential of the matter. In result good dispersing reduces the quantity of nanomaterials needed to achieve the same effects. As most nanomaterials are still fairly expensive, this aspect is of high importance for the commercialization of product formulations containing nanomaterials. Today, many nanomaterials are produced in a dry process. As a result, the particles need to be mixed into liquid formulations. This is where most nanoparticles form agglomerates during the wetting. Especially carbon nanotubes are very cohesive making it difficult to disperse them into liquids, such as water, ethanol, oil, polymer or epoxy resin. Therefore effective means of deagglomerating and dispersing are required to overcome the bonding forces after wettening the micron-powder or nano-powder.

## 4. EFFECTS

Ultrasound is used in a wide range of physical, chemical and biological processes. Homogenizing, emulsifying, and dispersing are examples for physical processes. Most of the applications of high-intensity ultrasound are based on cavitational effects. The physical effects of cavitation are being used in a top-down generation of nano-particles. Here, particles are reduced in size by the forces of cavitation. This includes the breaking of agglomerates and aggregates. The physical effects are used in combination with the chemical effects (sonochemistry) in the bottom-up production of nano-particles and crystals, i.e. during precipitation or crystallization. Here, ultrasound serves a number of roles in the initiation of seeding and subsequent crystal formation and growth.

### 4.1 Dispersing and Deagglomeration

Dispersion by ultrasound is a consequence of microturbulences caused by fluctuation of pressure and cavitation. Investigations at different materials, such as aqueous solutions of nanoparticulate siliciumdioxid powder and spray frozen agglomerates with a variable solid content have demonstrated the considerable advantage of ultrasound when compared with other technologies, such as rotor-stator mixers (e.g. turrax) or colloid mills. In particular for small matter from several nanometers to couple of microns, ultrasonic cavitation is very effective in breaking agglomerates, aggregates and even primaries. When ultrasound is being used for the milling of high concentration batches, the liquid jets streams resulting from ultrasonic cavitation, make the particles collide with each other at velocities of up to 1000km/h. This breaks van der Waals forces in agglomerates and even primary particles. Large particles are subject to surface erosion (via cavitation collapse in the surrounding liquid) or particle size reduction (due to fission through interparticle collision or the collapse of cavitation bubbles formed on the surface).

### 4.2 Emulsifying

If a cavitation bubble implodes near the phase boundary of two immiscible liquids the resultant shock wave can provide a very efficient mixing. Stable emulsions produced by sonication can be used in the textile, cosmetic, pharmaceutical, food, and petrochemical industry. Ultrasonically generated emulsions are often more stable and require less, if any surfactant than those produced conventionally. Since ultrasound is fully controllable and adaptable by the choice of amplitude, pressure and temperature, sonication is an effective instrument to obtain emulsions with smaller droplet sizes within a narrow size distribution.

### 4.3 Sonochemistry

Recent studies on the effect of sonication on suspended powders have shown that the particles can be forced into violent collision that, in the case of metals, fusion can occur. In some cases the colliding particles undergo chemical reactions. Thus, when copper and sulphur are sonicated together in hexane for 1 hour, 65% $Cu_2S$ is generated [1]. The chemical effects include the formation of $OH^-$ and $H^+$ species and hydrogen peroxide. The ultrasonic effects to chemical reactions – sonochemistry – include hydrolysis, oxidation, and depolymerization processes, too.

## 5. PROCESSING

Ultrasound describes a wide range technology. From cleaning tanks to sensors, from submersible transducers to nebulizers, from dental tools to baby images. These applications differ from each other because the parameters of the ultrasound differ a lot. For each specific application and process there is a specific optimal parameter configuration. For new processes and formulations this configuration has to be identified. In the following the parameters important to the liquid processing are examined.

## 5.1 Parameters

Ultrasonic liquid processing is described by a number of parameters. Most important are amplitude, pressure, temperature, viscosity, and concentration. The process result, such as particle size, for a given parameter configuration is a function of the energy per processed volume. The function changes with alterations in individual parameters. Furthermore, the actual power output per surface area of the sonotrode of an ultrasonic unit depends on the parameters.

### 5.1.1 Surface intensity

The power output per surface area of the sonotrode is the surface intensity (I). The surface intensity depends on the amplitude (A), pressure (p), the reactor volume ($V_R$), the temperature (T), viscosity ($\eta$) and others.

$$I[W/mm^2] = f(A[\mu m]_+, p[bar]_+, V_R[ml]_-, T[°C]_-, \eta[cP]_+, ...)$$

The impact of the generated cavitation depends on the surface intensity. So does the process result. The total power output of an ultrasonic unit is the product of surface intensity (I) and surface area (S):

$$P_L[W] = I[W/mm^2] * S[mm^2]$$

### 5.1.2 Amplitude

The amplitude of oscillation describes the way the sonotrode surface travels in a given time (e.g. 1/20,000s at 20kHz). The larger the amplitude, the higher is the rate at which the pressure lowers and increases at each stroke. In addition to that, the volume displacement of each stroke increases resulting in a larger cavitation volume (bubble size and/or number). When applied to dispersions, higher amplitudes show a higher destructiveness to solid particles. Table 1 shows general values for some ultrasonic processes.

| Process | Amplitude |
|---|---|
| Cleaning | 0.5 to 2 micron |
| Intensive Cleaning | 10 to 20 micron |
| Dispersing/Deagglomeration | 10 to 30 micron |
| Emulsifying | 20 to 60 micron |
| Primary Particle Reduction | 40 to 120 micron |

**Table 1 - General Recommendations for Amplitudes**

### 5.1.3 Pressure

Elevated pressure allows cavitation at temperatures close to or above the boiling point. It also increases the intensity of the implosion, which is related to the difference between the static pressure and the vapor pressure inside the bubble [5]. Since the ultrasonic power and intensity changes quickly with changes in pressure, a constant-pressure pump is preferable. When supplying liquid to a flow-cell the pump should be capable of handling the specific liquid flow at suitable pressures. Diaphragm or membrane pumps; flexible-tube, hose or squeeze pumps; peristaltic pumps; or piston or plunger pump will create alternating pressure fluctuations. Centrifugal pumps, gear pumps, spiral pumps, and progressive cavity pumps that supply the liquid to be sonicated at a continuously stable pressure are preferred.

### 5.1.4 Power and Intensity vs. Energy

Surface intensity and total power do only describe the intensity of processing. The sonicated sample volume and the time of exposure at a certain intensity have to be considered to describe a sonication process in order to make it scalable and reproducible. For a given parameter configuration the process result, e.g. particle size or chemical conversion, will depend on the energy per volume (E/V).

$$Result = f(E/V)$$

Where the energy (E) is the product of the power output (P) and the time of exposure (t).

$$E[Ws] = P[W] * t[s]$$

Changes in the parameter configuration will change the result function. This in turn will vary the amount of energy (E) required for a given sample value (V) to obtain a specific result value. For this reason it is not enough to deploy a certain power of ultrasound to a process to get a result. A more sophisticated approach is required to identify the power required and the parameter configuration at which the power should be put into the process material.

## 6. THREE STEP APPROACH

By dividing the process development into three stages, the costs and risks in the development can be reduced. Therefore the three steps are introduced: Feasibility Study, Optimization and Scale-Up. Because the sample volume is kept low, the development process can be speeded up while generating reproducible and scalable results based on the energy/volume.

## 6.1 Feasibility Study

The objective of the feasibility study is the identification of the potential for a specific product formulation or process. Unless there is data available from similar formulations and processes, this incorporates a trial-and-error testing of various parameter configurations.

When starting with a random parameter configuration, the likelihood to be at the optimal configuration is fairly low. Also the energy per volume required to obtain a result at this random or the optimal configuration is still to be discovered. Based on the premises that the process result for a given parameter configuration is a function of the energy per processed volume, if any result can be obtained, it will be better for a higher energy input. For this reason, the first tests should use a very high energy per volume level at high intensity levels. In this stage, trials using small sample volumes are preferred to shorten the time using a given equipment size. Lower volumes reduce the sample amount, too. This can be important for more precious materials. Most important is the even distribution of the sonication energy in the sample. Because the sonic intensity decreases with distance from the emitting surface, short distances are preferred. When a small liquid volume is exposed, the distance from the sonotrode can be kept short.

Table 2 shows typical energy/volume levels for sonication processes when being optimized. Since the first trials are not at optimum configuration, sonication at 10 to 50 times the typical value will show if there is any effect to the sonicated material or not.

| Process | Energy/ Volume | Sample Volume | Power | Time |
|---|---|---|---|---|
| Simple | < 100Ws/ml | 10ml | 50W | <20s |
| Medium | 100Ws/ml to 500Ws/ml | 10ml | 50W | 20 to 100s |
| Hard | > 500Ws/ml | 10ml | 50W | >100s |

**Table 2 - Typical Sonication Values**

As energy is transmitted into the liquid, the liquid will heat up. In order to allow for the input of high energy/volume levels cooling is required. Picture 1 shows a simple setup for the conduct of the feasibility trials. A cold water (or other coolant) bath is used to dissipate the heat generated in the small sample tube (green).

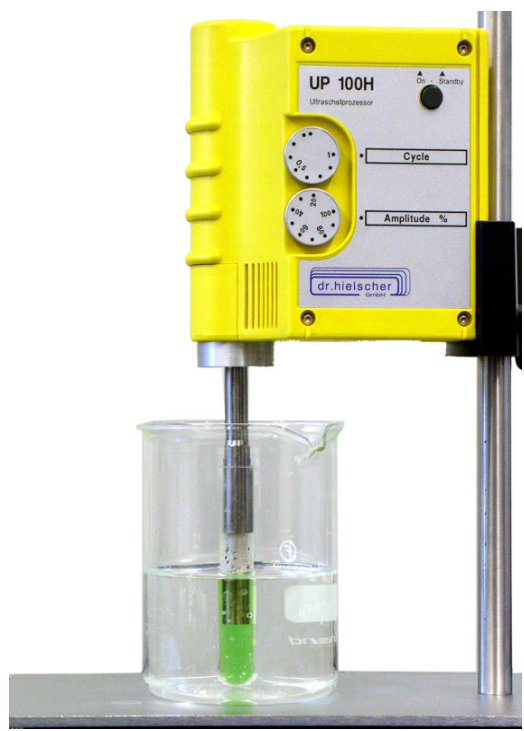

**Picture 1 – 100W Setup for Feasibility Testing**

By recording the actual power input, via a PC-interface or powermeter, the time of sonication and the actual amplitude and temperature in combination with the results of each trial a bottom line for the energy/volume can be established. If by chance an optimal configuration has been chosen, this configuration performance could be verified during optimization and could be scaled up to commercial level.

During feasibility study, the limits of sonication, e.g. temperature, amplitude or energy/volume for specific formulations should be examined, too. As ultrasound could generate negative effects to chemicals or particles, the critical levels for each parameter need to be examined in order to limit the following optimization to the parameter range where the negative effects are not observed.

For the feasibility study small lab or bench-top units are recommended to limit the expenses for equipment and samples in such trials. Generally 100 to 1,000 Watts units serve the purposes of the feasibility study very well.

## 6.2 Optimization

The results achieved during the feasibility study may show high energy/volume levels which when scaled up may be not cost effective. As these trials were to show the general impact of ultrasound only, optimization is needed to evaluate the cost effectiveness of ultrasound for the process. To identify the most energy and cost efficient way to deploy the ultrasonic energy in order to obtain the potential benefits observed in the feasibility study, a systematic optimization is required. In this step, the correlations between the relevant parameters – in particular amplitude, pressure and liquid formulation – are investigated. The objective is to find the optimal parameter configuration in which a minimum of energy is utilized to obtain the required benefit. This is the most efficient configuration. As this energy per volume value is the basis for the scale up to commercial level, efficiency gains from thorough optimization result in savings in operation and maintenance cost.

In this stage, systematic batch samples may help to identify the most important parameters for the specific process. Batch trials may also serve as indicators for the correlation between the specific parameter and the process result. Picture 2 shows the results of trials using different intensity and mass concentration values. The results indicate, that a higher Aerosil concentration improves the process efficiency.

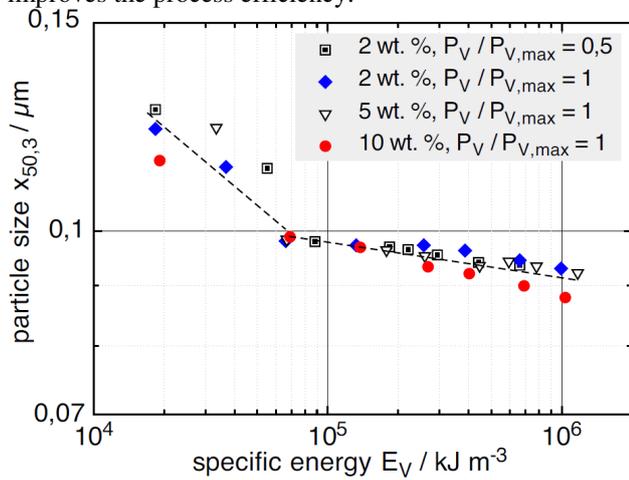

**Picture 2 - Influence of solid concentration on dispersion (UIP1000) [3]**

In a batch testing as shown in Picture 1 the pressure is limited to ambient pressure. For this reason, flow cells are required to determine the correlation between pressure and process efficiency.

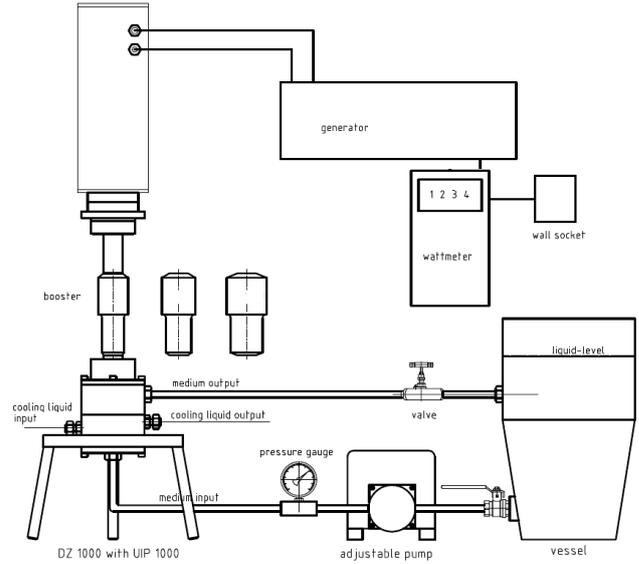

**Picture 3 - Flow Chart for Optimization Setup**

For the optimization, versatile ultrasonic units of 500 to 2,000 watts are suitable for this stage. In particular, if there is no information available on the sonication of similar materials the systems should be capable of covering a wide range for each parameter. For nanomaterials amplitudes from 10 to 100 micron should be tested. For most applications pressures from ambient to 2 or 3 bars above ambient are sufficient. For strong materials higher pressures of up to 50 bars above ambient may show better performance.

Again, the actual power input, the time of sonication, and the amplitude, pressure and temperature used are recorded together with the results of each trial. In the end the trial using the lowest energy per volume achieving a satisfying result will be chosen for scale up. It may be recommended to validate this particular trial by a second run at the same configuration.

### 6.3 Scale Up

Ultrasound is reproducible. When ultrasound is applied to an identical liquid formulation at an identical processing parameter configuration, the same energy per volume is required to obtain an identical result independent of the scale of processing. This allows for a linear scale up of the optimized parameter configuration to the full commercial scale.

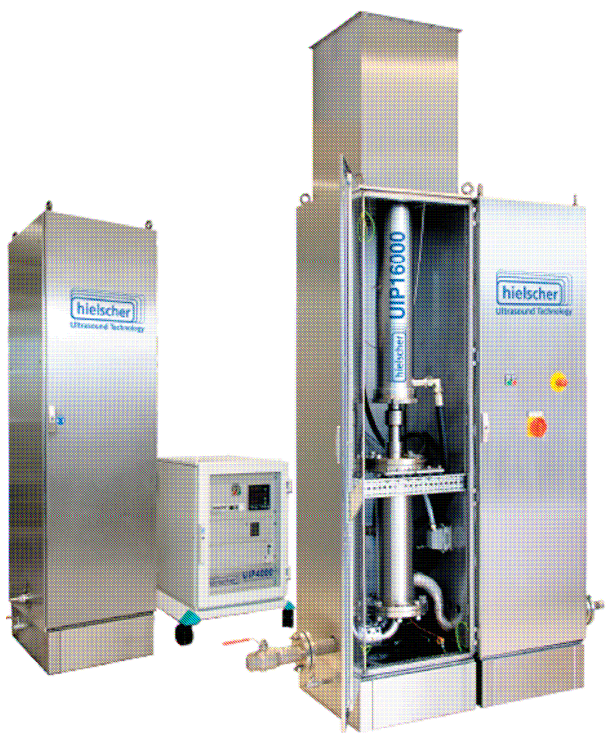

**Picture 4 - Commercial 4,000 Watts (left) and 16,000 Watts (right) System for Inline Operation**

Based upon the optimal parameter configuration found during optimization and the related energy/volume required the power capacity required for scale up can be calculated. The parameter configuration of the final system is to be identical to the optimal configuration. Commercial systems of up to 16,000 watts per single device are available. Such systems can be widely adapted to amplitude and pressure requirements in order to perform at the optimal efficiency. Picture 4 shows a complete 16,000 watts with generator, transducer, sonotrode and flow cell. For larger industrial setups, several such units can be used in parallel or in series. Such clusters are almost unlimited in total power. As the adaptation of such ultrasonic units is mostly limited to boosters, sonotrodes and flow cells, it mostly possible to re-adapt the units to new parameter configurations if necessary. This my be required, if the product formulation or the required result changes.

## 7. CONCLUSIONS

By a systematic approach to the development of ultrasonic processes, the costs and risks involved in the development and commercialization can be reduced. Because feasibility and optimization tests can be done in small lab or bench-top scale requiring small to medium size equipment, the time and expenses for this systematic identification of the optimal parameter configuration are low compared to the savings when scaling up to the production level. This approach does also help to find the important parameters behind the process result. This in turn can be used to obtain better results and a high reproducibility.